\documentclass[preprint,preprintnumbers,amsmath,amssymb]{revtex4}

\usepackage{graphicx}% Include figure files
\usepackage{dcolumn}% Align table columns on decimal point
\usepackage{bm}% bold math
\usepackage{amsmath}
\usepackage{xcolor}
\usepackage{comment}

\begin{document}

\title{\boldmath Qualitative Shadow Anomaly as a Smoking-gun Signature of Nonmetricity}

\author{Zi-Chao Lin$^a$\footnote{linzch24@cqu.edu.cn}}

\author{Hao Yu$^a$\footnote{yuhaocd@cqu.edu.cn}}

\author{Yu-Xiao Liu$^b$$^c$\footnote{liuyx@lzu.edu.cn}}

\author{Jin Li$^a$,\footnote{cqujinli1983@cqu.edu.cn (orresponding author)}}

\affiliation{$^{a}$Physics Department, Chongqing University, Chongqing 401331, China\\$^{b}$Lanzhou Center for Theoretical Physics, Key Laboratory of Theoretical Physics of Gansu Province, Key Laboratory of Quantum Theory and Applications of MoE, Gansu Provincial Research Center for Basic Disciplines of Quantum Physics, Lanzhou University, Lanzhou 730000, China$^{c}$\\Institute of Theoretical Physics $\&$ Research Center of Gravitation, School of Physical Science and Technology, Lanzhou University, Lanzhou 730000, China}

\begin{abstract}
	In general relativity and its metric-formalism extensions, the innermost stable circular orbit and the photon ring of a charged black hole contract as the charge increases. In this paper, we discover a qualitative anomaly that violates this universal behavior, serving as a smoking-gun signature of geometric nonmetricity. By considering a minimal coupling between the bosonic field and the independent affine metric, we demonstrate that nonmetricity induces an effective geometric force that triggers a trajectory expansion of both massless and massive bosons with increasing black hole charge. This qualitative inversion directly imprints onto the black hole shadow, lifting the degeneracy between metric-affine gravity and general relativity. Using Eddington-inspired Born-Infeld gravity as a concrete implementation, we show that next-generation horizon-scale imaging can resolve this signature, probing a complementary high-energy window up to the $10^{5}_{~}\,\text{TeV}$ scale.
\end{abstract}

\maketitle

\section{Introduction}\label{sec:intro}
The independence between the metric tensor and the affine connection is a foundational question in gravitational theories. In general relativity (GR), the equivalence principle identifies gravity with the Levi-Civita connection of spacetime. This standard metric formalism reduces the geometric degrees of freedom, thereby restricting the dynamics of the gravitational sector. Consequently, the metric-formalism extensions of GR that introduce higher-order curvature invariants will generate high-order derivatives in the field equations, triggering ghost instabilities~\cite{Stelle:1977ry,Woodard:2015zca}. Conversely, formulating gravity within the metric-affine framework offers a self-consistent and ghost-free alternative by treating the spacetime metric $g_{\mu\nu}^{~}$ and the connection $\Gamma^{\alpha}_{\mu\nu}$ as independent entities~\cite{Olmo:2011uz,Ferraris:1982wci,Vollick:2003qp}. A representative metric-affine framework is Eddington-inspired Born-Infeld (EiBI) gravity~\cite{Vollick:2003qp,Banados:2010ix}, where a characteristic scale $M_{\text{BI}}^{~}$ regularizes ultraviolet singularities by introducing Born-Infeld corrections before reaching the quantum gravity regime.

In the Einstein frame, the theory preserves the Einstein-Hilbert action with respect to the affine metric $q_{\mu\nu}^{~}$, while leaving matter fields non-minimally coupled via a constant $\bar{M}_{\text{BI}}^{~}=\sqrt{M_{\text{BI}}^{~}M_{\text{Pl}}^{~}}$~\cite{BeltranJimenez:2017doy}. This non-minimal coupling vanishes if the affine connection is determined solely by the spacetime metric $g_{\mu\nu}^{~}$, thus offering a window into the underlying geometric structure by probing effects beyond the scale $\bar{M}_{\text{BI}}^{~}$. Anomalous particle interactions beyond the Large Hadron Collider (LHC) threshold~\cite{ATLAS:2024vdo} and non-linear matter density modifications prior to big bang nucleosynthesis~\cite{Comelli:2004qr,Banados:2010ix} then establish constraints of $\bar{M}_{\text{BI}}^{~}\gtrsim10\,\text{TeV}$ and $\bar{M}_{\text{BI}}^{~}\gtrsim10^{-5}_{~}\,\text{TeV}$~\cite{Avelino:2012ge}, respectively. Therefore, seeking the Born-Infeld effects in the high-curvature regime beyond this scale is crucial. The manifestation of these effects around charged black holes~\cite{Vollick:2003qp,Banados:2010ix,Wei:2014dka,Shafeeque:2024agp} indicates that the recent horizon-scale imaging by the Event Horizon Telescope (EHT)~\cite{EventHorizonTelescope:2019dse,EventHorizonTelescope:2019ths,EventHorizonTelescope:2019ggy,EventHorizonTelescope:2022wkp,EventHorizonTelescope:2022wok} provides a powerful approach for testing this geometry. However, isolating the observation characteristics of the metric-affine structure remains a challenge. Under the traditional assumption, matter fields are minimally coupled to the spacetime metric $g_{\mu\nu}^{~}$. Therefore, the affine connection can only affect observables indirectly through the modified spacetime metric~\cite{Wei:2014dka,Yang:2021chw,Olmo:2023lil}. In the high-curvature regime of compact objects, such metric-induced corrections typically manifest as minor quantitative shifts in the black hole shadow radius. This degeneracy makes it difficult to distinguish the geometric signatures of nonmetricity from standard GR variations or matter-sector contributions.

Given that the spacetime metric is compatible with the Levi-Civita component of the affine connection, a fully metric-affine description of particle motion should be incorporated into the framework. In this paper, we break this degeneracy by considering the minimal coupling between the bosonic field and the affine metric $q_{\mu\nu}^{~}$ rather than the spacetime metric $g_{\mu\nu}^{~}$, implying that bosons are parallel-transported with respect to the independent affine connection. Physically, this coupling modifies the Killing vector, acting as an effective geometrical force on the particles. We demonstrate that it alters the bosonic geodesics around a charged black hole in an anomalous manner. Contrary to GR and its metric-formalism extensions, where the innermost stable circular orbit (ISCO) and the photon ring contract as the black hole accumulates charge, the nonmetricity triggers a universal trajectory expansion with increasing charge. This distinct qualitative inversion serves as a smoking-gun signature of geometric nonmetricity that directly imprints onto the optical appearance of the black hole shadow. By utilizing the profiles of 12 most promising supermassive black hole candidates~\cite{Zhang:2024owe}, we map the parameter space required for the EHT, the next-generation EHT (ngEHT)~\cite{Doeleman:2023kzg,Johnson:2023ynn}, and the Black Hole Explorer (BHEX)~\cite{Johnson:2024ttr} to resolve this signature. For a high-energy completion of gravity, we finally provide a prior estimate of the scale $\bar{M}_{\text{BI}}^{~}$, establishing a criterion to identify the affine structure.

\section{Affine-induced geodesic corrections}
Our starting point is a charged black hole in EiBI gravity with a cosmological constant $\Lambda=(\kappa-1)M_{\text{BI}}^{~}$. Assuming the matter field $\psi$ is minimally coupled to the spacetime metric, the model admits an analytical solution within the integration formalism~\cite{Banados:2010ix,Wei:2014dka}.  The solution is derived from the following action in the Born-Infeld frame~\cite{Banados:2010ix}
\begin{equation}
	S_{\text{EiBI}}^{~}
	=
	M_{\text{BI}}^{2}M_{\text{Pl}}^{2}\int\text{d}^{4}_{~}x\Big(\sqrt{-\det[g_{\mu\nu}^{~}+R_{\mu\nu}^{~}(\Gamma)/M_{\text{BI}}^{2}]}
	-\kappa\sqrt{-\det(g_{\mu\nu}^{~})}\,\Big)+S_{M}^{~}(\psi,g),
\end{equation}
where $\Gamma$ is the torsion-free affine connection treated independently of the spacetime metric. Departing from the initial version, we adopt a minimal coupling between the matter field and the affine metric $q_{\mu\nu}^{~}(g_{\mu\nu}^{~},\Gamma_{\mu\nu}^{\alpha})$ in the Born-Infeld regime. We assume that the original background solution~\cite{Banados:2010ix,Wei:2014dka} remains valid up to order $\sim\epsilon=Q^{2}_{~}/(M_{\text{BI}}^{2}r^{4}_{~})$. The expansion of the analytical solution shows that all higher-order corrections in terms of $\epsilon$ vanish when the radial distance $r$ and the scale $M_{\text{BI}}^{~}$ are large compared to the black hole charge $Q$. These higher-order terms correspond to curvature corrections to the Einstein-Hilbert action in the $M_{\text{BI}}^{2}\gg|R_{\mu\nu}^{~}|$ regime. The coupling between $M_{\text{BI}}^{~}$ and $Q$ in the solution highlights that the nonmetricity of the theory is sensitive to the matter sector. At infinity, the metrics coincide with their standard GR form ($q_{\mu\nu}^{(0)}=g_{\mu\nu}^{(0)}$), leaving only the Levi-Civita component of the affine connection non-vanishing. Consequently, the theory effectively recovers GR through local metric compatibility.

In this charged black hole background, the motion of a null boson with four-momentum $p^{\mu}_{~}=\text{d}x^{\mu}_{~}/\text{d}\lambda$ is governed by the geodesic equation $p^{\alpha}_{~}\nabla_{\alpha}^{(\Gamma)}p_{\mu}^{~}=0$ compatible with $q_{\mu\nu}$. This equation has an equivalent form
\begin{equation}\label{GQ1}
	\frac{\text{d}p_{\mu}^{~}}{\text{d}\lambda}=\Big(\frac{1}{2}\partial_{\mu}^{~}q_{\alpha\beta}^{~}-\frac{1}{M_{\text{BI}}^{2}}\partial_{(\alpha}^{~}R_{\beta)\mu}^{~}+\frac{1}{M_{\text{BI}}^{2}}R_{\mu\rho}^{~}\Gamma_{\alpha\beta}^{\rho}\Big)p^{\alpha}_{~}p^{\beta}_{~},
\end{equation}
where the right-hand side represents a correction to the standard  geodesic equation of GR, acting as an effective geometric force on the boson. This affine-induced geometric force breaks the universality of free fall, as it originates from the nonmetricity scaled by $M_{\text{BI}}^{-2}$ and vanishes asymptotically at spatial infinity where $q_{\mu\nu}$ reduces to $g_{\mu\nu}$. Eq.~\eqref{GQ1} indicates that a specific component $p_{\kappa}^{~}$ is conserved if the corresponding right-hand side vanishes. Defining a Killing vector associated with this conserved quantity via $K_{\alpha}^{~}p^{\alpha}_{~}=p_{\kappa}^{~}$, where the index $\kappa$ is a fixed label denoting the symmetric direction and does not enter into standard index summation, one finds $p^{\alpha}_{~}\nabla_{\alpha}^{(\Gamma)}p_{\kappa}^{~}=p^{\alpha}_{~}p^{\beta}_{~}(\partial_{(\alpha}^{~}K_{\beta)}^{~}-\Gamma_{\alpha\beta}^{\rho}K_{\rho}^{~})$, representing the Killing equation within the metric-affine framework. Together with Eq.~\eqref{GQ1}, this implies that the Killing vector is fully determined by the affine metric. We expand the Killing vector as $K_{\mu}^{~}=K_{\mu}^{(0)}+K_{\mu}^{(1)}+\mathcal{O}(\epsilon^{2}_{~})$. The leading-order term depends solely on the isometry of the standard metric, $q_{\mu\nu}^{(0)}$ or $g_{\mu\nu}^{(0)}$, of the black hole. We can therefore choose $K_{~}^{(0)}=\partial_{\kappa}^{~}$ (or $K_{\kappa}^{(0),\mu}=\delta_{\kappa}^{\mu}$) to isolate a specific direction $x^{\kappa}_{~}$ of which the affine metric is independent. It is straightforward to verify that the leading-order expression of Eq.~\eqref{GQ1} yields $\partial_{\kappa}^{~}q^{(0)}_{\mu\nu}=0$, and that $K_{\mu}^{(0)}$ satisfies the Killing equation at this order.

The first-order correction $K_{\mu}^{(1)}$ is governed by the linearized Killing equation,
\begin{equation}\label{KE1}
	q_{\alpha(\mu}^{(0)}\partial_{\nu)}^{~^{~}_{~}}K_{~}^{\alpha,(1)}+\frac{1}{2}K_{~}^{\alpha,(1)}\partial_{\alpha}^{~}q_{\mu\nu}^{(0)}-\Gamma_{\mu\nu}^{\kappa,(1)}g_{\kappa\kappa}^{(0)}=0,
\end{equation}
where the second term vanishes due to the background isometry, provided the first-order correction takes the form $K_{~}^{(1),\mu}=\delta^{\mu}_{\kappa}f(x^{\mu}_{~})$. In this case, the function $f(x^{\mu}_{~})$ is uniquely determined by the nonmetricity component of the affine connection at order $\sim\epsilon$ via the relation $\delta^{\kappa}_{(\mu}\partial_{\nu)}^{~^{~}_{~}}f=\Gamma_{\mu\nu}^{\kappa,(1)}$. The underlying geometric structure requires
\begin{equation}
	\frac{1}{2}\partial_{\kappa}^{~}q_{\mu\nu}^{(1)}-\partial_{(\mu}^{~^{~}_{~}}q_{\nu)\kappa}^{(1)}+\partial_{(\mu}^{~^{~}_{~}}g_{\nu)\kappa}^{(1)}+(q_{\alpha\kappa}^{(1)}+g_{\alpha\kappa}^{(1)})\Gamma^{\alpha,(0)}_{\mu\nu}=0.
\end{equation}
This expression is derived by expanding Eq.~\eqref{GQ1} in powers of $\epsilon$, where corrections scaled by $M_{\text{BI}}^{~}$ are mapped to $q^{(1)}_{~}$ and $g^{(1)}_{~}$ using the field equations from Ref.~\cite{Banados:2010ix}. Utilizing the identity $q_{\nu\alpha}^{~}(q^{-1}_{~})^{\mu\alpha}_{~}=\delta^{\mu}_{\nu}$ alongside Eq.~\eqref{KE1}, we arrive at the constraint $\partial_{\mu}^{~}f=\partial_{\mu}^{~}[g^{(1)}_{\kappa\kappa}/q^{(0)}_{\kappa\kappa}]$. This is a universal property of metric-affine theories when the affine structure is torsion-free and symmetric. For a charged black hole of mass $M$, the Killing vector $K_{\mu}^{t}$ associated with the conserved quantity $p_{t}^{~}$ is given by
\begin{equation}
	f_{t}^{~}(r,\theta,\phi)=f_{t}^{~}(r)=-\frac{C}{5}\frac{Q^{2}_{~}+5 r (r - M)}{Q^{2}_{~}+r (r - 2 M)}\frac{Q^{2}_{~}}{M_{\text{BI}}^{2}r^{4}_{~}},
\end{equation}
where the integration constant has been omitted since the theory recovers local metricity at spatial infinity. The dimensionless constant $C\sim \bar{M}_{\text{BI}}^{~}/M_{*}^{~}$ acts as a universal parameter weighting the coupling strength between the bosonic field and the affine metric, where $M_{*}^{~}\gtrsim M_{\text{BI}}^{~}$ denotes the characteristic mass scale of the coupling. Conversely, for the Killing vector $K_{\mu}^{\phi}$, the constraint reduces to $q_{\phi\phi}^{(0)}\partial_{\mu}^{~}f_{\phi}^{~}=0$. To satisfy the asymptotic boundary conditions of the metrics, we set $f_{\phi}^{~}(t,r,\theta)=0$.

\section{Boson trajectory inversion and anomaly}
Within the framework of EIBI gravity, the causal structure of the matter field is not uniquely determined by the spacetime metric. Consequently, the motion of null bosons around the charged black hole is modified by the underlying nonmetricity of the theory. The effect is characterized by the first-order correction $K^{(1)}_{\mu}$ to the Killing vector. For a boson restricted to a constant polar angle $\theta=\theta_{0}^{~}$, the radial motion obeys
\begin{equation}\label{TE1}
	\frac{1}{p_{\phi}^{2}}\Big(\frac{\text{d}r}{\text{d}\lambda}\Big)^{2}_{~}=-\frac{1}{g_{tt}^{~}}\frac{1}{g_{rr}^{~}}\frac{1}{K_{t}^{2}}\frac{1}{b^{2}_{~}}-\frac{1}{g_{rr}^{~}}\frac{1}{r^{2}_{~}\sin^{2}_{~}\theta_{0}^{~}}.
\end{equation}
At leading order, it reduces to the standard trajectory equations of GR. The higher-order corrections become significant when the curvature scale $|R_{\mu\nu}^{~}|$ approaches $\bar{M}_{\text{BI}}^{2}$ near the black hole. At  order $\sim\epsilon$, the radial motion can be rewritten as $(\text{d}r'/\text{d}\lambda)^{2}_{~}=V_{\text{eff}}^{(1)}(r')$ by introducing the coordinate transformation $r\rightarrow r'=\int\sqrt{-g_{tt}^{~}g_{rr}^{~}}K_{t}^{~}\text{d}r$. This redefinition decouples the correction $V_{\text{eff}}^{(1)}$ to the effective potential $V_{\text{eff}}^{~}(r')=V_{\text{eff}}^{~}(r)=g_{tt}^{~}K_{t}^{2}/(r^{2}_{~}\sin^{2}_{~}\theta_{0}^{~})$ from the impact parameter $b$. Consequently, the radius $r_{0}^{~}$ of the circular orbit depends solely on the spacetime geometry. In a model-independent form, corrections to the effective potential can be generally decomposed as $V_{\text{eff}}^{(1)}\propto g_{tt}^{(1)}(1+1/q_{tt}^{(0)})$, where the latter term explicitly parameterizes the coupling between the boson field and the independent affine geometry. Since $q_{tt}^{(0)}<0$ outside the event horizon, its contribution introduces a universal repulsive potential that counteracts the metric-sector attraction. In the context of EiBI gravity, this generic feature is explicitly governed by the screening parameter $C$. As the mass scale $M_{*}^{~}$ descends from the Planck scale toward $\bar{M}_{\text{BI}}^{~}$, the first-order correction to $K_{\mu}^{t}$ becomes dominant. Hence, $C$ serves as a screening parameter that quantifies the deviation of the effective potential from the form derived in Ref.~\cite{Olmo:2023lil}. Since the coordinate transformation reduces to an identity mapping at leading order, $V_{\text{eff}}^{(0)}$ and the relevant physical quantities naturally recover the standard results in GR.

In the new coordinates ($t,r',\theta,\phi$), the circular orbit of the boson is defined by the conditions $\text{d}r'/\text{d}\lambda=0$ and $\text{d}^{2}_{~}r'/\text{d}\lambda^{2}_{~}=0$. The radius and impact parameter of the orbit are determined by the equivalent conditions $V_{\text{eff}}^{~}|_{r'_{0}}=1/b^{2}_{0}$ and $\partial_{r'}^{~}V_{\text{eff}}^{~}|_{r'_{0}}=0$. Focusing on unstable orbits, we impose the additional constraint $\partial_{r'}^{2}V_{\text{eff}}^{~}|_{r'_{0}}<0$ to numerically obtain the orbital radius $r_{0}^{~}$ and the corresponding critical impact parameter $b_{0}^{~}$.
\begin{figure}[!htb]
\center{
\includegraphics[width=7cm]{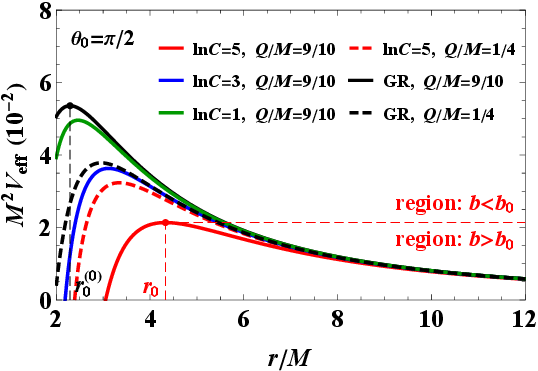}
}
\caption{Radius $r_{0}^{~}$ of the unstable circular orbit for a null boson coupled to the affine structure with $M_{\text{BI}}^{~}=1/M$. For comparison, the black curves denote the radii  $r_{0}^{(0)}$ predicted by GR.}
\label{NuBos}
\end{figure}
As illustrated in Fig.~\ref{NuBos}, the deviation of the effective potential from the standard GR profile near the black hole is sensitive to the parameter $C$. In the limit $C\to 0$, the residual deviation arises solely from the corrections to the spacetime metric. In this case, the effective potential reduces to the form presented in Ref.~\cite{Olmo:2023lil}, consistent with a purely metric-induced modification. The deviation is also modulated by the black hole charge $Q$. For a fixed value of $C$, the divergence between the EiBI prediction $r^{~}_{0}$ and the GR prediction $r^{(0)}_{0}$ becomes more conspicuous as the black hole approaches the extremal limit. Conversely, for weakly charged black holes, distinguishing the effects of nonmetricity on the unstable circular orbit radius becomes observationally challenging.

A remarkable feature shown in Fig.~\ref{NuBos} is that the orbital radius increases with the black hole charge when $C$ is non-vanishing. This anomalous behavior undergoes an inversion than GR, where increasing the black hole charge contracts the photon sphere due to the enhanced gravitational pull of the mass-energy localized in the Maxwell field. This trajectory expansion occurs because the affine-induced term $f_{t}^{~}(r)$ acts as an effective repulsive geometric barrier in the high-curvature regime, overpowering the metric-sector attraction. This qualitative anomaly is universal and persists for massive bosons. For a massive boson, the radial dynamics around the charged black hole at a fixed polar angle $\theta=\theta_{0}^{~}$ is governed by Eq.~\eqref{TE1} together with a correction term $s/(g_{rr}^{~}p_{\phi}^{2})$, where $s$ is an unitary parameter with dimensions of length squared. As before, the circular orbit of the massive boson satisfies analogous conditions on the redefined effective potential $\tilde{V}_{\text{eff}}^{~}(r')=\tilde{V}_{\text{eff}}^{~}(r)=g_{tt}^{~}K_{t}^{2}[1/(r^{2}_{~}\sin^{2}_{~}\theta_{0}^{~})+s/p_{\phi}^{2}]$. Here, we numerically determined the radius $r_{p}^{~}$ of the stable orbit with the condition $\partial_{r'}^{2}\tilde{V}_{\text{eff}}^{~}|_{r'_{p}}>0$.
\begin{figure}[!htb]
\center{
\includegraphics[width=7cm]{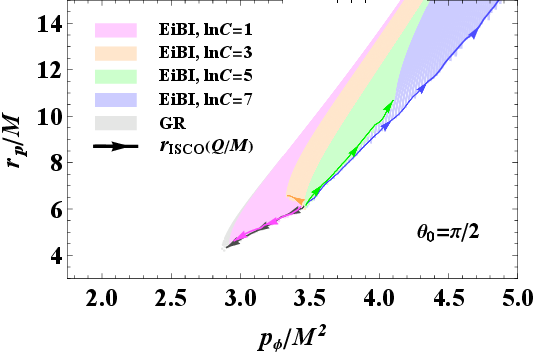}
}
\caption{Dependence of the ISCO radius for a massive boson on the black hole charge $Q$ (solid line) with $M_{\text{BI}}^{~}=1/M$. The rescaled charge $Q/M$ increases from 0 to 1 in the direction of the arrows while the screening parameter $C$ is held constant. The intersection of the curves indicates an attractor at $Q/M=0$. The shaded region denotes the parameter space for stable circular orbits.}
\label{MasBos}
\end{figure}
As demonstrated in Fig.~\ref{MasBos}, the non-minimal coupling between the boson and the affine metric also affects the stable orbital radius. For the massive boson in a specified state around the black hole with a given charge, the radius $r_{p}^{~}$ will increase monotonically with the screening parameter $C$. Consequently, the ISCO radius $r_{\text{ISCO}^{~}}\!\!$, which corresponds to the lower boundary of the shaded region, deviates from the GR prediction in the presence of a non-vanishing $C$. At the $Q/M=0$ attractor, this deviation vanishes, indicating that EiBI gravity recovers effective metric compatible. Crucially, for a charged black hole, the ISCO radius can increase as a function of $Q/M$. This behavior contrasts with GR, where the ISCO radius decreases with increasing $Q/M$. This universal trajectory anomaly provides a distinct smoking-gun signature of geometric nonmetricity. Our numerical results spanning the overall $M_{\text{BI}}^{~}M$ parameter space reveal that this feature emerges as the screening parameter reaches $C\sim3$.

\section{Optical signature and parameter constraint}
We next focus on the dynamics of photons around a charged black hole in EiBI gravity. Following standard conventions, emission trajectories are categorized into three classes based on the orbit number $n$, which counts the number of equatorial plane crossings: direct emission for $n<3/4$, lensing ring emission for $3/4<n <5/4$, and photon ring emission for $n>5/4$. Assuming photons are emitted from the north pole with a constant polar angle $\theta=\theta_{0}^{~}$, the orbit number satisfies $2\pi n=\phi_{\text{total}}^{~}$. Under the variable substitution $r\rightarrow u=1/r$, the total deflection angle is given by $\phi_{\text{total}}^{~}=2\int^{u_{c}^{~}}_{0}\Phi(u)\text{d}u$, where the integrand $\Phi(u)=\text{d}\phi/\text{d}u$ is derived from the trajectory equation,
\begin{equation}\label{TE3}
	\Big(\frac{\text{d}r}{\text{d}\phi}\Big)^{2}_{~}=-\frac{r^{2}_{~}\sin^{2}_{~}\theta_{0}^{~}}{g_{rr}^{~}}\Big(\frac{r^{2}_{~}\sin^{2}_{~}\theta_{0}^{~}}{g_{tt}^{~}K_{t}^{2}}\frac{1}{b^{2}_{~}}+1\Big).
\end{equation}
The upper integration limit $u_{c}^{~}$ is classified by the impact parameter. For an infalling photon ($b<b_{0}^{~}$), $u_{c}^{~}$ is given by the radius of the event horizon $r_{\text{BH}}^{~}$. For an unstable circular orbit ($b=b_{0}^{~}$), $u_{c}^{~}=1/r_{0}^{~}$. For an escaping photon ($b>b_{0}^{~}$), a turning point occurs at the effective potential barrier, meaning $u_{c}^{~}$ is the root of $1/\Phi(u)=0$. Assuming an optically and geometrically thin accretion disk in the equatorial plane as the emission source, we neglect radiative transfer and disk structural effects on the photon flux. The observed intensity $I_{\text{obs}}^{~}(b)$ is obtained by summing the local emission from each successive intersection with the accretion disk, $I_{\text{obs}}^{~}\propto\sum_{m}^{~}I_{\text{em}}^{~}|_{r=r_{m}^{~}(b)}$.
\begin{figure}[!htb]
\center{
\includegraphics[width=7cm]{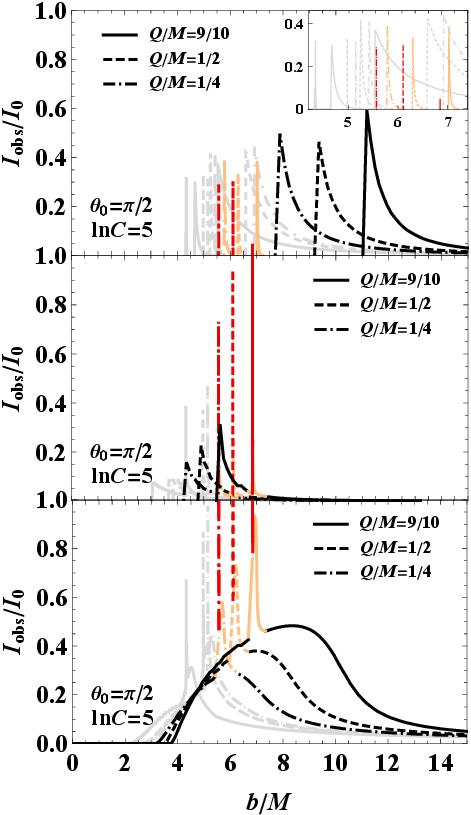}
}
\caption{Anomaly in the optical appearance of a charged black hole shadow as perceived by a distant observer. The observed intensity is normalized to the peak emitted intensity $I_{0}^{~}$. From top to bottom, the panels correspond to three emission models where the accretion disk extends to $r_{\text{ISCO}}^{~}$, $r_{0}^{~}$, and $r_{\text{BH}}^{~}$, respectively. Curves are categorized based on the photon orbit number $n$: the photon ring ($n>5/4$, red), the lensing ring ($3/4<n<5/4$, orange), and the direct emission ($n<3/4$, black). For comparison, the corresponding GR shadow profiles are indicated by gray curves.}
\label{OpticAppComb}
\end{figure}
Consequently, the photon ring emission manifests as a sharp peak in the observed intensity profile (see Fig.~\ref{OpticAppComb}), offering an observational approach to identify the anomalous charge-induced $r_{0}^{~}$ expansion in the black hole shadow. However, the direct emission contributes a broad background intensity due to its minimal intersection number and infinitely extended impact parameter range. Because of this exponential demagnification~\cite{Gralla:2019xty}, the localized brightness of the narrow photon ring risks being blurred within the integrated average brightness profile.

The source intensity $I_{\nu}^{~}(r,\nu)$ of the surrounding accretion flow dominates the optical appearance of the black hole. The total emitted intensity $I_{\text{em}}^{~}(r)$ is obtained by integrating the specific intensity over all frequencies. For photons traveling from the accretion disk to a distant observer, the frequency is gravitationally shifted to $\nu'=\sqrt{g_{tt}^{~}}\,\nu$. Because the phase-space density $I_{\nu}^{~}/\nu^{3}_{~}$ remains invariant along the geodesic~\cite{Lindquist:1966igj,Bromley:1996wb}, the observed intensity profile scales with the local emission as $g_{tt}^{2} I_{\text{em}}^{~}(r)$. In our analysis, we employ three distinct profiles for the emitted intensity $I_{\text{em}}^{~}$, as detailed in Ref.~\cite{Gralla:2019xty}. Numerical results, presented in Fig.~\ref{OpticAppComb}, demonstrate that the photon ring radius expands as the black hole accumulates charge. This trend represents a striking anomaly in the optical appearance of the black hole shadow. The direct emission exhibits analogous behavior, provided that the inner boundary scales with the black hole charge in a similar manner. This result is physically intuitive, as photon trajectories are strongly modified by the non-minimal coupling in the strong-field regime where $|R_{\mu\nu}^{~}|\gtrsim\bar{M}_{\text{BI}}^{2}$. Therefore, the innermost boundary of the average intensity profile can be employed to characterize the emergence of the anomaly.

\begin{figure}[!htb]
\center{
\includegraphics[width=7cm]{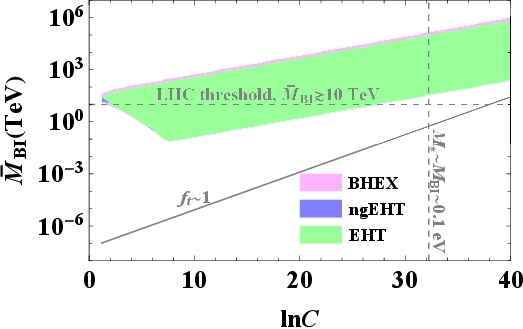}
}
\caption{Projected constraints on the universal coupling constant $\bar{M}_{\text{BI}}^{~}$ in EiBI gravity. Shaded regions represent the parameter space estimated from the 12 most promising observational candidates, while dashed lines denote the lower bounds derived from LHC constraints ($\bar{M}_{\text{BI}}^{~}\gtrsim10\,\text{TeV}$). The region above the solid line satisfies the expansion condition $f_{t}^{~}\lesssim1$.
}
\label{MBICons}
\end{figure}

The parameter constraints can be established by systematically scanning the entire parameter space to identify anomalous signatures in targeted observations. To determine optimistic constraints, we consider a typical emission model in which the accretion disk extends to the ISCO. We incorporate the physical parameters of 12 target black holes (see Table~1 in Ref.~\cite{Zhang:2024owe}), which are primary candidates for observations by the EHT, ngEHT, and BHEX, due to their large apparent shadow sizes. Our numerical results (see Fig.~\ref{MBICons}) suggest the potential for detecting shadow anomalies  within the range $10^{-2}\,\text{TeV}\lesssim\bar{M}_{\text{BI}}^{~}\lesssim10^{6}\,\text{TeV}$, with BHEX offering the most extensive parameter space coverage owing to its enhanced angular resolution. However, only a marginal, order-of-magnitude level improvement is observed in the BHEX constraints, primarily because the angular sizes of the ISCO for these candidates are comparable to the baseline resolutions of the EHT and ngEHT. The validity of our numerical results is verified by checking the expansion condition across the entire sample of 12 candidates. Furthermore, the lower boundary of the parameter space is determined by excluding regions where the ISCO fails to exist. This bound remains robust across various emission models, owing to the regularity of the metric solution within specific parameter regimes. The upper bound is then given by the limit $M_{*}^{~}\gtrsim M_{\text{BI}}^{~}\gtrsim 10^{-4}_{~}\,\text{TeV}$ on the screening parameter $C$. Notably, incorporating the complementary LHC constraint of $\bar{M}_{\text{BI}}^{~}\gtrsim10\,\text{TeV}$ refines our final joint parameter space to $10\,\text{TeV}\lesssim\bar{M}_{\text{BI}}^{~}\lesssim10^{5}\,\text{TeV}$, mapping out a wide discovery space for horizon-scale shadow imaging.

\section{Discussions}
An issue in testing charge-dependent modifications is that astrophysical black holes are expected to be rapidly neutralized by plasma. However, we emphasize that the charge $Q$ employed in our derivation serves as a mathematical proxy for any source field that effectively couples the matter sector to the affine connection. In generalized gravitational frameworks, identical field equation structures can emerge without actual electric charges. For instance, in braneworld scenarios, a tidal charge arises via the nonlocal projection of bulk spacetime onto the brane~\cite{Dadhich:2000am}. Alternatively, within the mimetic framework~\cite{Chaichian:2014qba}, an effective scalar-induced field can mimic the stress-energy tensor through conformal invariance~\cite{Chen:2017ify}. Therefore, the universal trajectory expansion discovered here stands as a robust model-independent signature of geometric nonmetricity, decoupling from the constraints of astrophysical plasma neutralization.

\section{Conclusion}
In this paper, we established that a minimal coupling between bosonic matter and the independent connection in metric-affine gravity generates a qualitative inversion of particle geodesics. The resulting expansion of the photon sphere and ISCO with increasing source charge contradicts the standard behavior of GR. This anomaly provides a smoking-gun signature of nonmetricity that bypasses metric degeneracies and maps onto the high-resolution intensity profiles of black hole shadows. The observation by the EHT, ngEHT, and BHEX opens a window to test the affine structure of spacetime up to the $10^{5}_{~}\,\text{TeV}$ scale.

\section*{Acknowledgements}
We acknowledge useful discussions with Yungui Gong, Shao-Qi Hou, Wu-Zhong Guo, and Qing-Hua Zhu. This work is supported by the National Key Research and Development Program of China (Grant No.~2021YFC2203004), the Fundamental Research Funds for the Central Universities Project (Grant No.~2024IAIS-ZD009), and the National Natural Science Foundation of China (Grants No.~12505057, No.~12547101, No.~12475056, and No.~12247101).

\end{document}